\documentclass[onecolumn,amsmath,amssymb,pra,superscriptaddress,a4paper,floatfix,showkeys,showpacs]{revtex4}
\usepackage{graphicx}
\usepackage{mathptmx}      
\begin{document}

\title{Multi-order interference is generally nonzero}

\author{Hans De Raedt}
\email{h.a.de.raedt@rug.nl}
\affiliation{%
Department of Applied Physics,
Zernike Institute for Advanced Materials,
University of Groningen, Nijenborgh 4, NL-9747 AG Groningen, The Netherlands
}%
\author{Kristel Michielsen}
\email{k.michielsen@fz-juelich.de}           
\affiliation{%
Institute for Advanced Simulation, J\"ulich Supercomputing Centre,
Research Centre J\"ulich, D-52425 J\"ulich, Germany
}%
\author{Karl Hess}
\email{k-hess@illinois.edu}           
\affiliation{%
Beckman Institute, Department of Electrical Engineering and Department
of Physics, University of Illinois, Urbana, Il 61801, USA
}%
\date{\today}

\begin{abstract}
It is demonstrated that the third-order interference, as obtained from explicit solutions of Maxwell's equations
for realistic models of three-slit devices, including an idealized version
of the three-slit device used in a recent three-slit experiment with light
(U. Sinha {et al.}, Science 329, 418 (2010)), is generally nonzero.
The hypothesis that the third-order interference
should be zero is shown to be fatally flawed
because it requires dropping the one-to-one correspondence
between the symbols in the mathematical theory and the different experimental configurations.
\end{abstract}

\maketitle

\section{Introduction}\label{Introduction}

According to the working hypothesis (WH) of Refs.~\cite{SORK94,FRAN10,SINH10},
quantum interference between many different
pathways is simply the sum of the effects from all pairs of pathways.
In particular, application of the WH to a three-slit experiment yields~\cite{SINH10}
\begin{eqnarray}
I(\mathbf{r},\mathrm{OOO})&=&|\psi_1(\mathbf{r})+\psi_2(\mathbf{r})+\psi_3(\mathbf{r})|^2
,
\label{three0}
\end{eqnarray}
where $\psi_j$ with $j=1,2,3$ represents the amplitude of the wave emanating from the $j$th slit with the other two slits closed
and $\mathbf{r}$ denotes the position in space.
Here and in the following, we denote the intensity of light recorded in a
three-slit experiment by $I(\mathbf{r},\mathrm{OOO})$, the triple O's indicating that
all three slits are open.
We will write $I(\mathbf{r},\mathrm{COO})$ for the intensity of light recorded in
the experiment in which the first slit is closed, and so on.

Assuming the WH to be correct, it follows that
\begin{eqnarray}
I(\mathbf{r},\mathrm{OOO})&=&|\psi_1(\mathbf{r})+\psi_2(\mathbf{r})+\psi_3(\mathbf{r})|^2
\nonumber \\
&=&|\psi_1(\mathbf{r})+\psi_2(\mathbf{r})|^2+|\psi_1(\mathbf{r})+\psi_3(\mathbf{r})|^2+|\psi_2(\mathbf{r})+\psi_3(\mathbf{r})|^2
\nonumber \\
&&-|\psi_1(\mathbf{r})|^2-|\psi_2(\mathbf{r})|^2-|\psi_3(\mathbf{r})|^2
\nonumber \\
&=&I(\mathbf{r},\mathrm{OOC})+I(\mathbf{r},\mathrm{OCO})+I(\mathbf{r},\mathrm{COO})
\nonumber \\&&
-I(\mathbf{r},\mathrm{OCC})-I(\mathbf{r},\mathrm{COC})-I(\mathbf{r},\mathrm{CCO})
.
\label{three1}
\end{eqnarray}
In other words, still assuming the WH to be correct, we must have
\begin{eqnarray}
\Delta(\mathbf{r})&=&I(\mathbf{r},\mathrm{OOO})-I(\mathbf{r},\mathrm{OOC})
-I(\mathbf{r},\mathrm{OCO})-I(\mathbf{r},\mathrm{COO})
\nonumber \\&&
+I(\mathbf{r},\mathrm{OCC})+I(\mathbf{r},\mathrm{COC})+I(\mathbf{r},\mathrm{CCO})=0
.
\label{three2}
\end{eqnarray}

According to Refs.~\onlinecite{SORK94,SINH10}, the identity Eq.~(\ref{three2})
follows from quantum theory and the assumption that the Born rule $I(\mathbf{r})\propto|\Psi(\mathbf{r})|^2$ holds.
In a recent three-slit experiment with light~\cite{SINH10}, the seven contributions
to $\Delta(\mathbf{r})$ were measured and taking into account the uncertainties intrinsic to
these experiments, it was found that $\Delta(\mathbf{r})\approx0$.
This finding was then taken as experimental evidence that the Born rule $I(\mathbf{r})\propto|\Psi(\mathbf{r})|^2$
is not violated~\cite{SINH10}.

The purpose of the present paper is to draw attention to the fact that
within Maxwell's theory or quantum theory,
the premise that Eq.~(\ref{three0}) (which implies Eq.~(\ref{three2})) holds is generally false.
By explicit solution of the Maxwell equations for several devices, including an idealized version
of the three-slit device used in experiment~\cite{SINH10}, we show that $\Delta(\mathbf{r})$
is generally nonzero.
We also point out that the reasoning that leads to the WH~\cite{SORK94,FRAN10,SINH10} and to the conclusion
that $\Delta(\mathbf{r})=0$ is fundamentally flawed because it does not satisfy one
of the basic criteria of a proper mathematical description of a collection of experiments, namely
that there should be a one-to-one correspondence between the symbols in the mathematical
theory and the different experimental configurations.

\section{Solution of Maxwell's equation}

A conclusive test of the WH Eq.~(\ref{three0}) is to simply solve
the Maxwell equations for a three-slit device in which slits can be opened or closed (simulation results
for the device employed in the experiment reported in Ref.\onlinecite{SINH10} are presented in Section~\ref{merit}).
For simplicity, we assume translational invariance in the direction along the long axis of the slits,
effectively reducing the dimension of the computational problem by one.

\begin{figure*}[t]
\begin{center}
\mbox{
\includegraphics[width=7cm ]{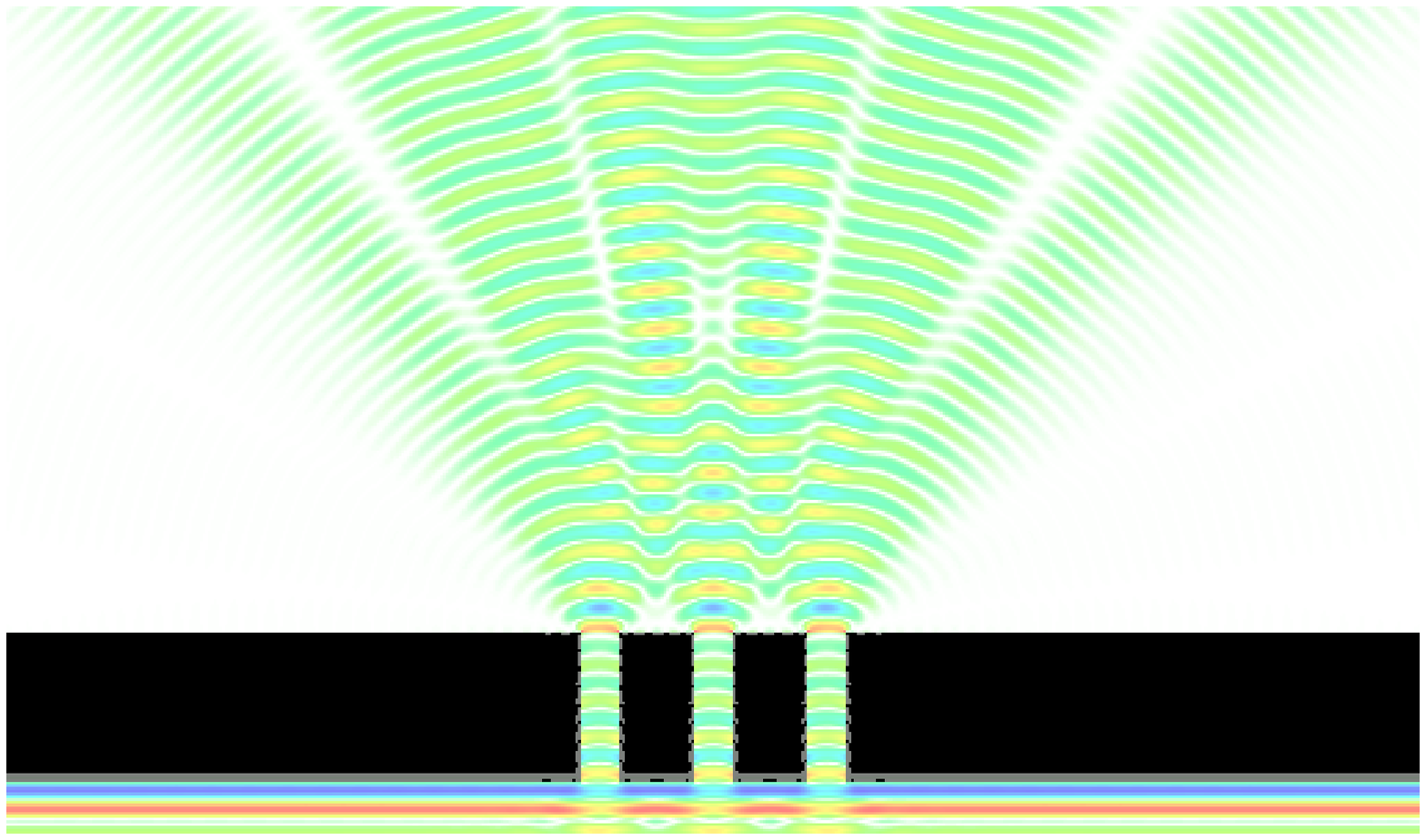}
\hbox to 0.5cm{}
\includegraphics[width=7cm ]{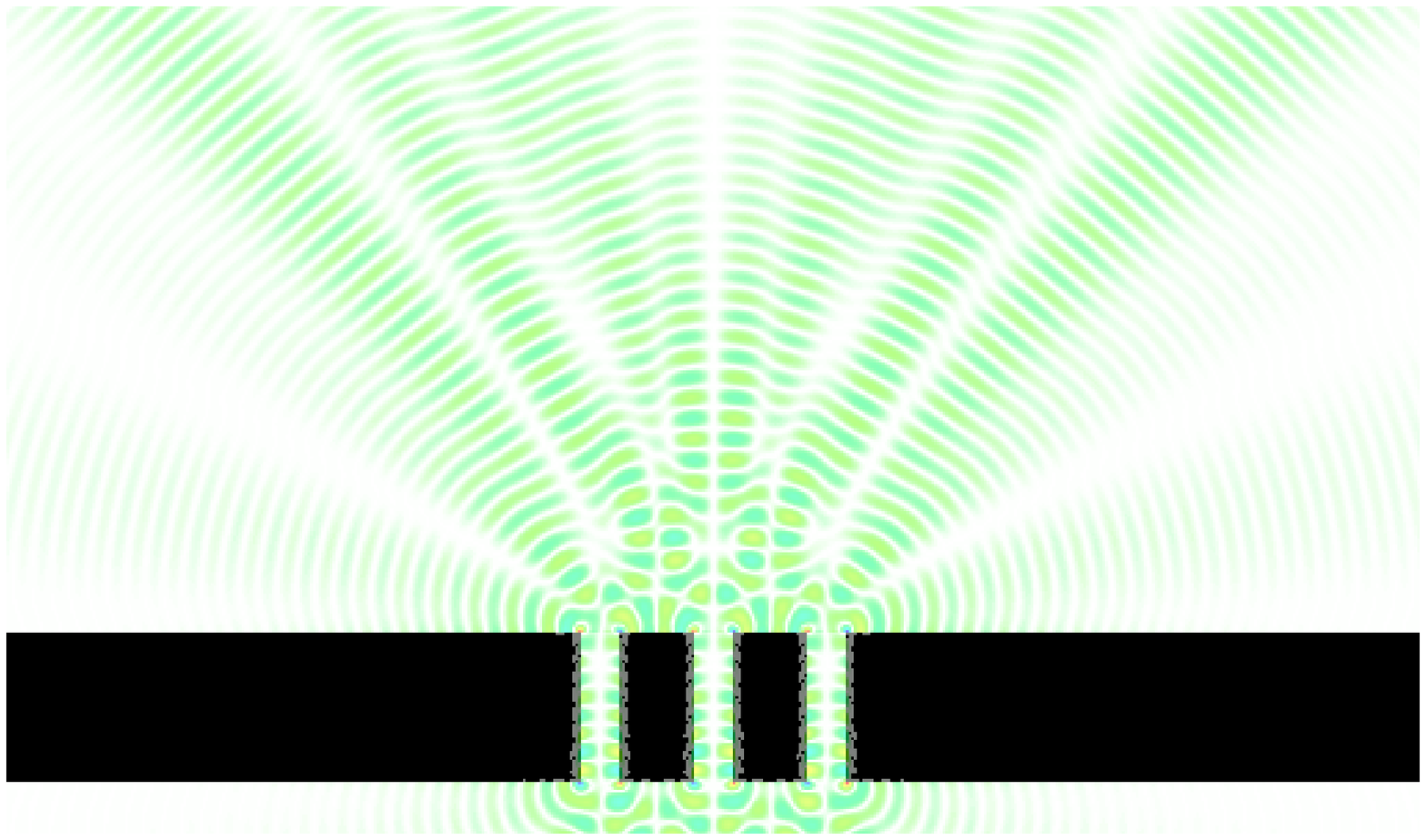}
}
\caption{%
Amplitudes of the $E_x$ (left) and $E_z$ (right) components of the electric fields as obtained from
a FDTD solution of Maxwell's equation for light incident on a metallic plate with three slits.
The incident wave is monochromatic and has wavelength $\lambda=500$nm.
The slits are $\lambda$ wide, their centres being separated by $3\lambda$.
The index of refraction of the $4\lambda$-thick metallic plate (colored black) is $2.29+2.61 i$.
In the FDTD simulations, the material (steel) is represented by a Drude model~\cite{TAFL05}.
}
\label{fig.lambda}
\end{center}
\end{figure*}

\begin{figure*}[t]
\begin{center}
\includegraphics[width=7cm ]{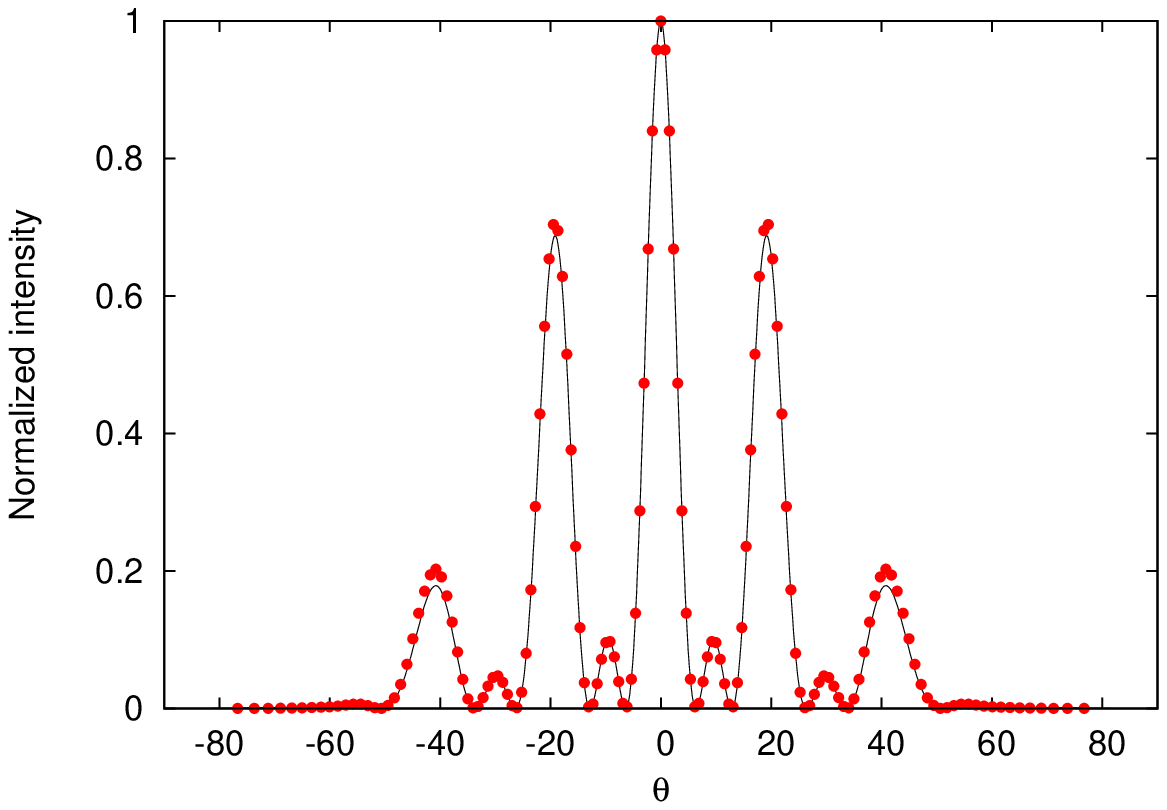}
\includegraphics[width=7cm ]{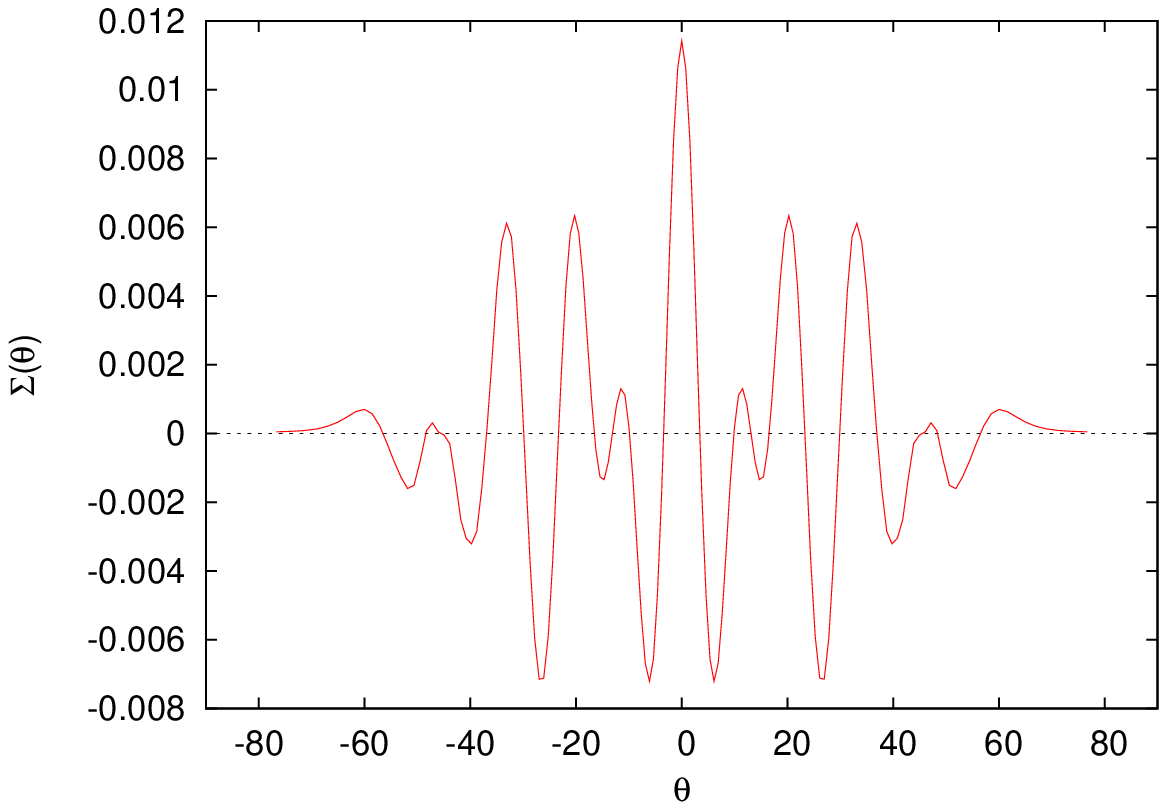}
\caption{%
Left:
Angular distribution of light $I(\theta;\mathrm{OOO})$ transmitted by $N=3$ slits (see Fig.~\ref{fig.lambda})
as obtained from the FDTD simulation (bullets) and Fraunhofer theory (solid line)
$I(\theta,s=1,d=2,N=3)$, see Eq.~(\ref{multi0}), where $s$ and $d$ are the dimensionless slit width and slit separation,
respectively~\cite{BORN64}.
Right:
Normalized difference
$\Sigma(\theta)=[I(\theta;\mathrm{OOO})-I(\theta;\mathrm{COO})-I(\theta;\mathrm{OCO})-I(\theta;\mathrm{OOC})+I(\theta;\mathrm{CCO})
+I(\theta;\mathrm{COC})+I(\theta;\mathrm{OCC})]/I(\theta=0;\mathrm{OOO})$
as a function of $\theta$.
According to the WH of Refs.~\cite{SORK94,FRAN10,SINH10}, this difference should be zero.
}
\label{fig.lambda.compare}
\label{fig.lambda.combine}
\end{center}
\end{figure*}

In Fig.~\ref{fig.lambda} we show the stationary-state solution of the Maxwell equations,
as obtained from a finite-difference time-domain (FDTD) simulation~\cite{TAFL05}.
From the simulation data, we extract the angular distribution $I(\theta,\mathrm{OOO})$.
Repeating these simulations with one and two of the slits closed,
we obtain $I(\theta,\mathrm{COO})$ and so on.
In all these simulations, the number of mesh points per wavelength $\lambda$ was taken to be 100 to ensure that
the discretization errors of the electromagnetic (EM) fields and geometry are negligible.
The simulation box is $75\lambda\times40\lambda$ large (corresponding to 30 011 501 grid points),
terminated by UPML boundaries
to suppress reflection from the boundaries~\cite{TAFL05}.
The device is illuminated from the bottom (Fig.~\ref{fig.lambda}), using a current source
that generates a monochromatic plane wave that propagates in the vertical direction.

In Fig.~\ref{fig.lambda.compare}(left) we show a comparison between the angular distribution
of the transmitted intensity $I(\theta;\mathrm{OOO})$ as obtained from the FDTD simulation (bullets)
and Fraunhofer theory (solid line).
Plotting
\begin{eqnarray}
\Sigma(\theta)&=&
\frac{I(\theta;\mathrm{OOO})-I(\theta;\mathrm{COO})-I(\theta;\mathrm{OCO})-I(\theta;\mathrm{OOC})+I(\theta;\mathrm{CCO})}{I(\theta=0;\mathrm{OOO})}
\nonumber \\&&
+\frac{I(\theta;\mathrm{COC})+I(\theta;\mathrm{OCC})}{I(\theta=0;\mathrm{OOO})}
,
\end{eqnarray}
as a function of $\theta$ (see Fig.~\ref{fig.lambda.compare}(right)) shows,
beyond any reasonable doubt,
that the WH Eq.~(\ref{three0}) of Refs.~\cite{SORK94,FRAN10,SINH10}
is in conflict with Maxwell's theory: $\Sigma(\theta)$ takes values in the 0.5\%
range, much too large to be disposed of as numerical noise.
Note that $\Sigma(\theta)$ is obtained from data produced by
seven different device configurations.

Physically, the fact that $\Sigma(\theta)\not=0$ is related to
the presence of wave amplitude in the vicinity
of the surfaces of the scattering object (one, two, or three slit system),
see for instance Fig.~\ref{fig.lambda}(right).
These amplitudes are very sensitive to changes in the geometry
of the device, in particular to the presence or absence of a sharp edge.
Although these amplitudes themselves do not significantly contribute
to the transmitted light in the forward direction, it is well-known
that their existence affects the transmission properties of the device
as a whole~\cite{GAY06,LALA06}.

\subsection{Illustrative example}

The essence of a wave theory is that the whole system is described by one, and only one, wave function.
Decomposing this wave function in various parts that are solutions
of other problems and/or to attach physical relevance to parts of the wave
is a potential source for incorrect conclusions and paradoxes.
Even for one-and-the-same problem, the idea to think in terms of waves
made up of other waves can lead to nonsensical conclusions,
such as that part of a light pulse can travel at a superluminal velocity.
Of course, we may express the wave field
as a superposition of a {\bf complete} set of basis functions,
e.g. by Fourier decomposition,
and this may be very useful to actually solve the mathematical problem
(to a good approximation).
However such decompositions are primarily convenient mathematical tricks which,
in view of the fact that in principle any complete set of basis functions could be used,
should not be over-interpreted as being physically relevant~\cite{ROYC10,ROYC10b}.

The WH Eq.~(\ref{three0}) takes these ideas substantially further
by decomposing the wave amplitude in three parts,
each part describing the same system (a single slit)
located at a different position in space.
It is then conjectured that the amplitude for the whole system (three slits)
is just the sum of these three different amplitudes.

Advocates of the ``physical'' motivation for this conjecture might
appeal to Feynman's path integral formulation~\cite{FEYN65b} of wave mechanics
to justify their picture but in fact,
one can see immediately from Feynman's path integral formalism that the WH Eq.~(\ref{three0}) is not valid.

We use the expression for the propagator of the electron as given by Feynman and assume that the particle proceeds from a location $a$ and
time $t_a$ on one side of the screen with slits labeled $1, 2, 3$ to a location $b$ where a measurement is taken at time $t_b$ on the
other side. We assume that there exists some time $t_c$ between $t_a$ and $t_b$
(as assumed by Feynman on p. 36, Ref.~\onlinecite{FEYN65b}).
The propagator for this process is denoted by Feynman as $K(b, a)$~\cite{FEYN65b}.
If we include for clarity the times then we would have to write $K((b, t_b), (a, t_a))$.
As pointed out by Feynman (Ref.~\onlinecite{FEYN65b}, p. 57)
we have a connection of this propagator to the wave function $\psi$ given by:
\begin{equation}
\psi(b, t_b) = K((b, t_b), (a, t_a)).
\label{nov22n1}
\end{equation}
Feynman represented the propagator $K$ by a path integral
that sums over all possible space-time paths to go from $a$ to $b$ with the
end-point times as given above.
If we have an infinitely extended screen in between $a, b$ with only slit $1$ open, then all
paths can only proceed through this one slit. We denote the wave function that is calculated for a path leading through a
particular point $x_1$ of the slit at time $t_{x_1}$ by $\psi_1'$.
Similarly for slits $2$ and $3$ open only we have $\psi_2'$ and $\psi_3'$
respectively and the corresponding $K$'s are calculated with Feynman paths that only go through slits $2$ or $3$
respectively.

Had we chosen all three slits open, then Feynman's formalism
insists that pathways going through multiple slits matter in general.
Therefore, we would have to include paths through multiple slits in the path
integral representation of $K$ and we would obtain a corresponding $\psi_{123}'$.
Thus, Feynman's quantum mechanics with all three slits open does contain an infinity of
paths that go through multiple slits resulting in $\psi_{123}'$.
However, none of the wave functions $\psi_1'$, $\psi_2'$ or $\psi_3'$
may contain any path through more than one slit because of the assumption that only one slit be open at a time.
Therefore all the expressions involving these amplitudes do not contain
multiple-slit path integrals and consequently do not contain all the paths that
are required to compute $\psi_{123}'$.
We illustrate the importance of ``{\bf all}'' by solving the Maxwell equations
for a minor variation of the three-slit experiment in which we block one slit.

\begin{figure*}[t]
\begin{center}
\includegraphics[width=7cm ]{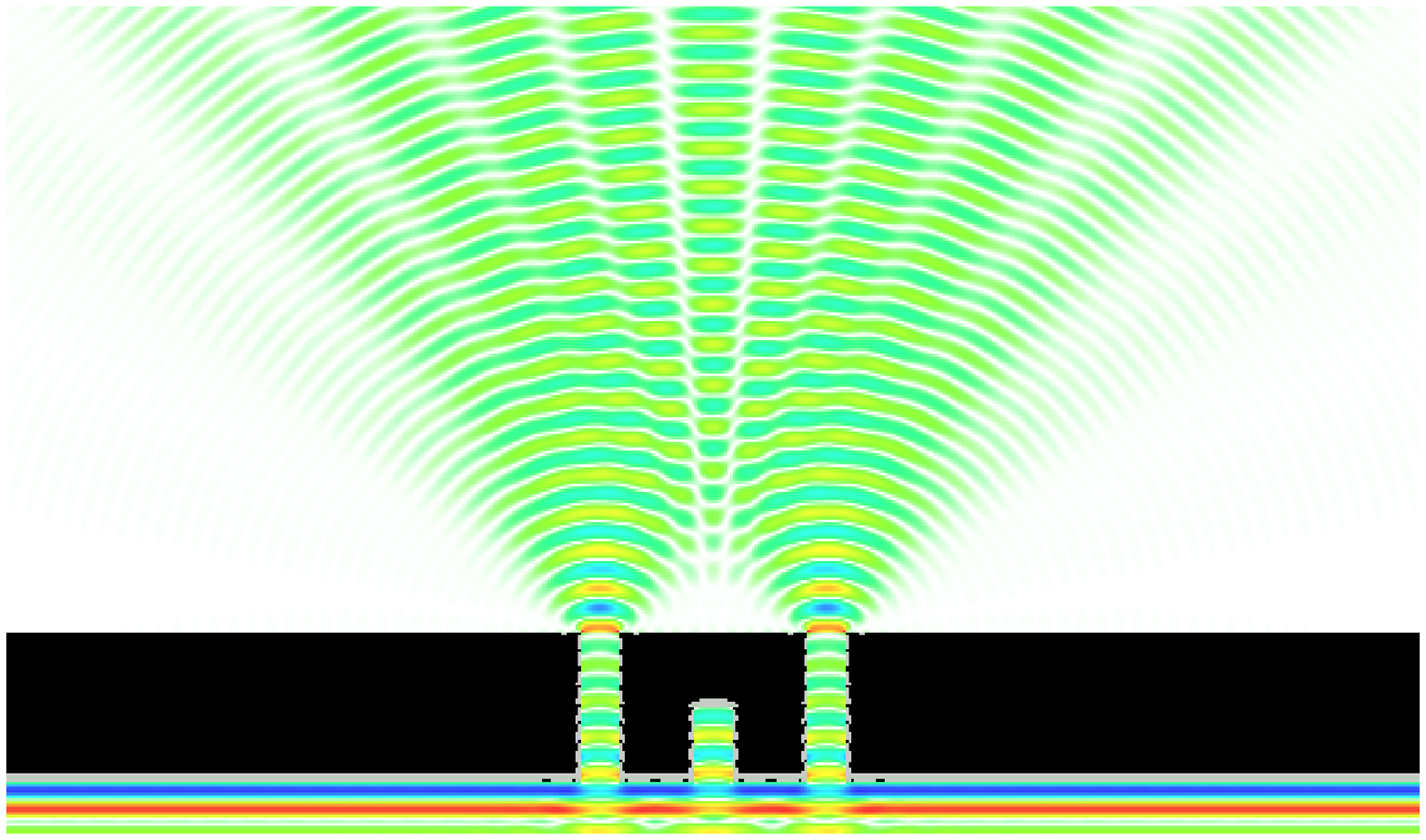}
\includegraphics[width=7cm ]{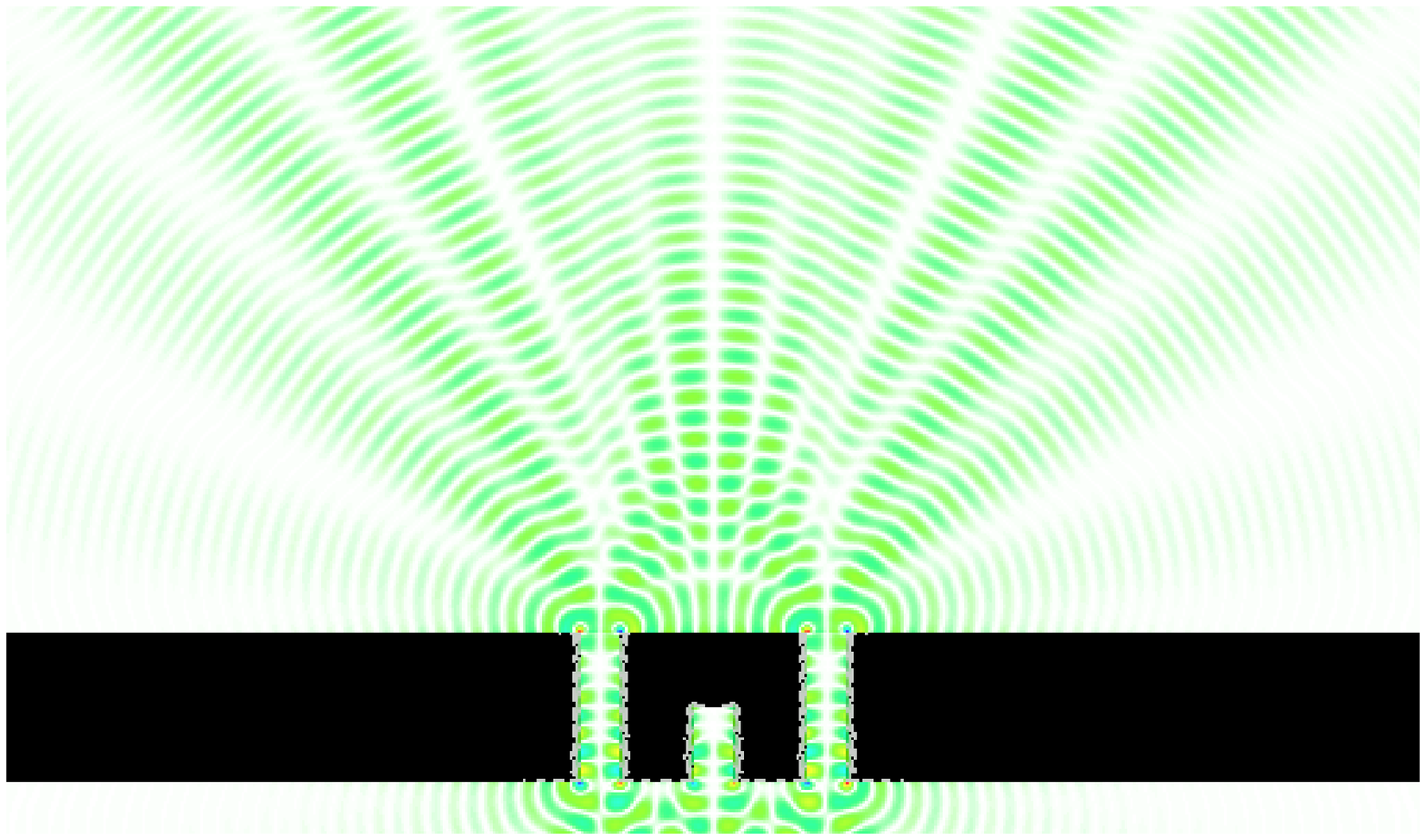}
\includegraphics[width=7cm ]{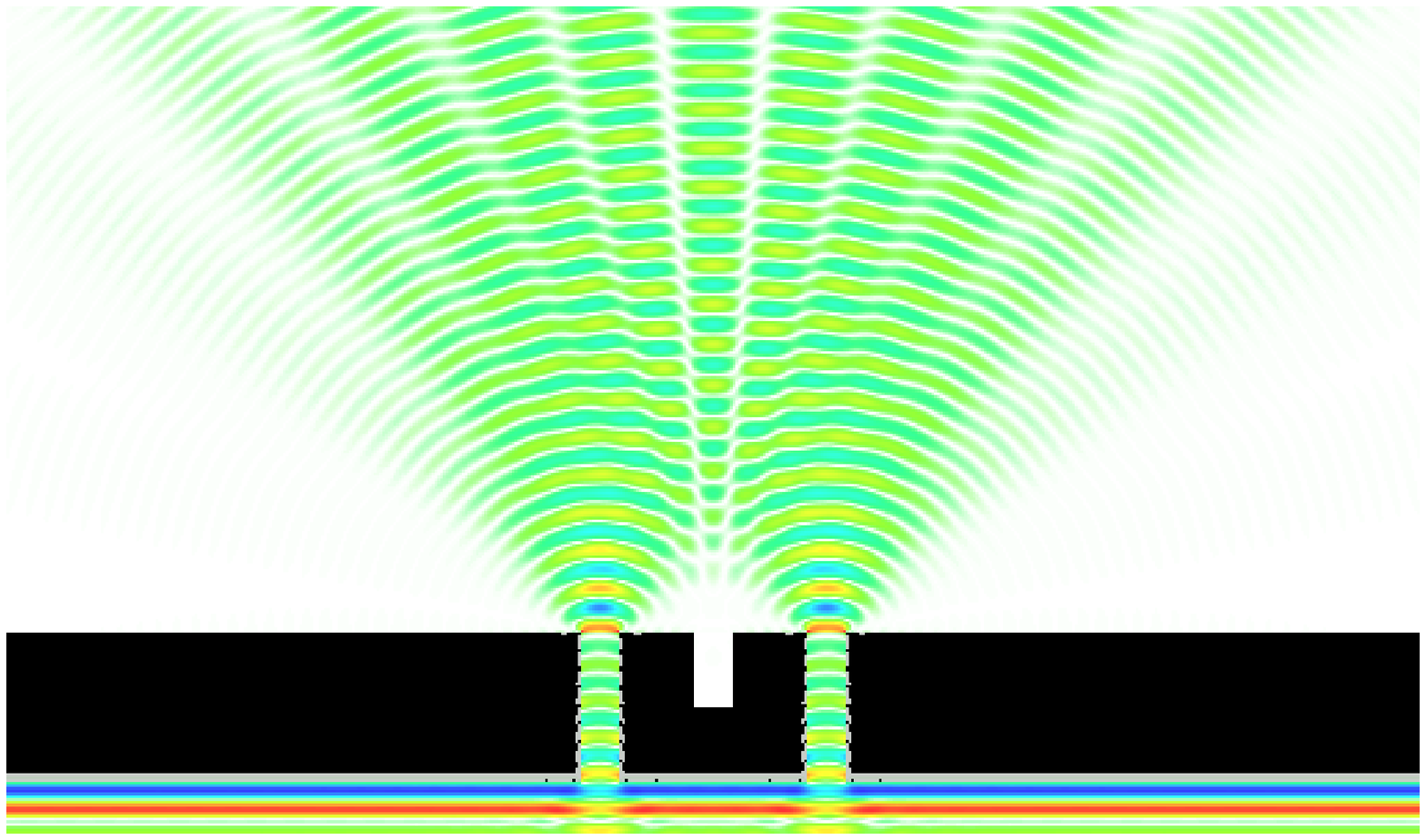}
\includegraphics[width=7cm ]{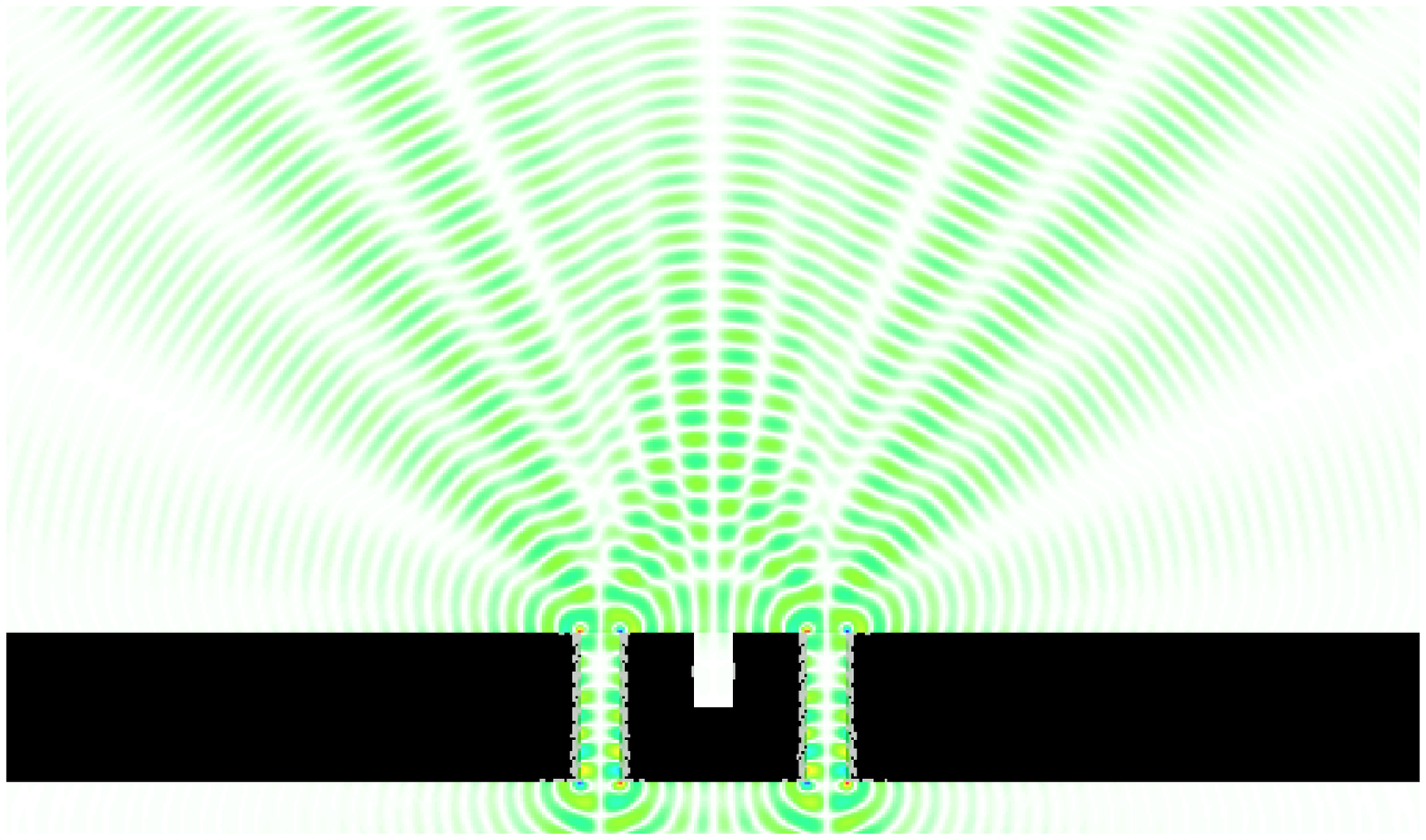}
\includegraphics[width=7cm ]{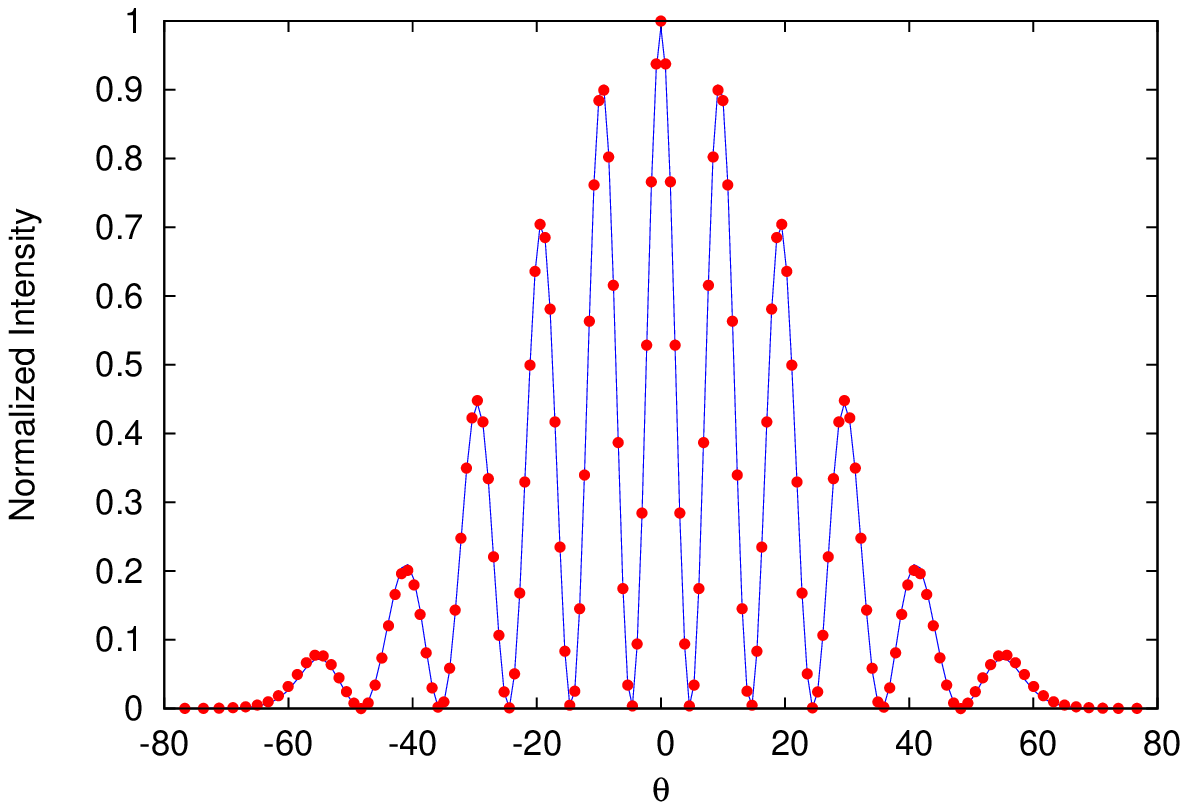}
\includegraphics[width=7cm ]{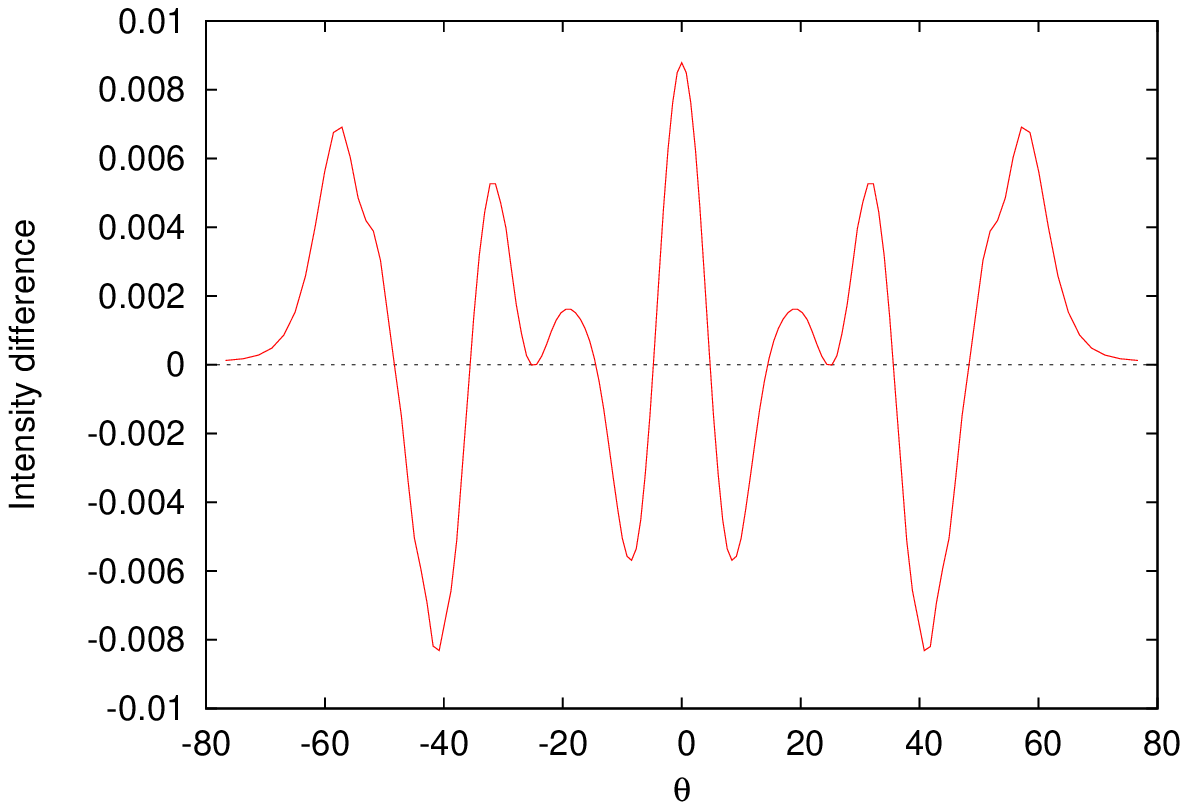}
\caption{%
Top and middle:
Amplitudes of the $E_x$ (left) and $E_z$ (right) components of the electric fields as obtained from
a FDTD solution of Maxwell's equation for light incident on a metallic plate with two slits
and a hole between the two slits.
The incident wave is monochromatic and has wavelength $\lambda=500$nm.
The slits are $\lambda$ wide, their centres being separated by $3\lambda$.
The index of refraction of the $4\lambda$-thick metallic plate (steel, colored black) is $2.29+2.61 i$.
The holes are $\lambda$ wide and $2\lambda$ deep.
Bottom left:
Angular distribution of light transmitted by the devices extracted from the FDTD simulation data.
Solid line: Slit in the centre filled with material half-way from the top (see top panel);
Red bullets: Slit in the centre filled with material half-way from the bottom (see center panel).
On the scale used, the two angular distributions cannot be distinguished.
Right:
Normalized difference between the angular distributions
of the device with the hole in the bottom (top panel)
and the top (middle panel).
Adopting the reasoning of Refs.~\cite{SORK94,FRAN10,SINH10}, this difference should be zero.
}
\label{fig.half2}
\end{center}
\end{figure*}

The geometry of the device that we consider is depicted in Fig.~\ref{fig.half2},
together with the FDTD solution of the EM fields in the stationary state.
We have taken the three-slit device and blocked the middle slit
by filling half of this slit with material (the same as used for other parts
of the three-slit device), once from the top, Fig.~\ref{fig.half2}(top),
and once from the bottom, Fig.~\ref{fig.half2}(middle).
Comparing the FDTD solutions shown in
Fig.~\ref{fig.half2}(top) and Fig.~\ref{fig.half2}(middle),
it is obvious to the eye that the wave amplitudes
provide no support for the idea that these
systems can be described by a wave going through one slit
and another wave going through the other slit.
As in the case of the three slits, the angular distributions
for the two cases look very similar (see Fig.~\ref{fig.half2}(bottom,left))
but differ on the one-percent level.

\section{When does the working hypothesis hold?}\label{whholds}

It is of interest to scrutinize the situations for which the WH Eq.~(\ref{three0}) is correct~\cite{KHRE08a,UDUD10}.
As pointed out earlier, in general, interference between many different
pathways is {\bf not} simply the sum of the effects from
all pairs of pathways. To establish nontrivial conditions under which it truly is a pairwise sum,
we discard experiments for which the WH trivially holds, that is we discard
experiments that {\it exactly} probe the interference of three waves,
such as the extended Mach-Zehnder interferometer experiment described in Ref.~\onlinecite{FRAN10}
and the class of statistical problems described by trichotomous variables
considered in Ref.~\onlinecite{NYMA11}.

Let us (1) neglect the vector character of EM waves
and (2) assume that the diffraction of the three-slit system is described
by Fraunhofer diffraction theory.
Then, for normal incidence, the angular distribution of
light intensity produced by diffraction from $N$ slits is given by~\cite{BORN64}
\begin{eqnarray}
I(\theta,s,d,N)&=&
\left(\frac{\sin(N\pi d\sin\theta)}{\sin(\pi d\sin\theta)}\right)^2
\left(\frac{\sin(\pi s\sin\theta)}{\pi s\sin\theta}\right)^2
,
\label{multi0}
\end{eqnarray}
where $s$ and $d$ are the dimensionless slit width and slit separation
expressed in units of the wavelength $\lambda$, respectively.
Therefore, we have
\begin{eqnarray}
\Delta(\theta)&=&I(\theta,s,d,3)-2I(\theta,s,d,2)-I(\theta,s,2d,2)+3I(\theta,s,d,1)
\nonumber\\
&=&
\left[(1+2\cos 2ad)^2-8\cos^2 ad-4\cos^22ad+3\right]
\left(\frac{\sin as}{as}\right)^2
=0
,
\end{eqnarray}
where $a=\pi \sin\theta$.
Thus, in the Fraunhofer regime the WH Eq.~(\ref{three0}) holds.

It is not difficult to see that $\Delta(\theta)=0$
is an accident rather than a general result
by simply writing down the Maxwell curl equations~\cite{BORN64,TAFL05}
\begin{eqnarray}
\epsilon(\mathbf{r})\frac{\partial \mathbf{E}(\mathbf{r},t)}{\partial t}
&=&\nabla \times \mathbf{H}(\mathbf{r},t)-J(\mathbf{r},t)
\nonumber \\
\mu(\mathbf{r})\frac{\partial \mathbf{H}(\mathbf{r},t)}{\partial t}
&=&\nabla \times \mathbf{E}(\mathbf{r},t)
,
\label{maxwell}
\end{eqnarray}
where the geometry of the device is accounted for by the permittivity $\epsilon(\mathbf{r})$
and for simplicity, as is often done in optics~\cite{BORN64}, we may assume that
the permeability $\mu(\mathbf{r})=1$.

Let us write $\epsilon(\mathbf{r},\mathrm{OOO})$ for the permittivity of the three-slit geometry
and $\mathbf{E}(\mathbf{r},t,\mathrm{OOO})$, $\mathbf{H}(\mathbf{r},t,\mathrm{OOO})$ for the corresponding solution
of the Maxwell equations Eq.~(\ref{maxwell}).
The WH Eq.~(\ref{three0}) asserts that
there should be a relation between
($\mathbf{E}(\mathbf{r},t,\mathrm{OOO})$, $\mathbf{H}(\mathbf{r},t,\mathrm{OOO})$)
and
($\mathbf{E}(\mathbf{r},t,\mathrm{COO})$, $\mathbf{H}(\mathbf{r},t,\mathrm{COO})$),
($\mathbf{E}(\mathbf{r},t,\mathrm{OCO})$, $\mathbf{H}(\mathbf{r},t,\mathrm{OCO})$),
...,
($\mathbf{E}(\mathbf{r},t,\mathrm{OCC})$, $\mathbf{H}(\mathbf{r},t,\mathrm{OCC})$)
but this assertion is absurd:
There is no theorem in Maxwell's theory that relates the solutions for the case
$\epsilon(\mathbf{r},\mathrm{OOO})$
to solutions for the cases
$\epsilon(\mathbf{r},\mathrm{COO})$...
$\epsilon(\mathbf{r},\mathrm{OCC})$.
The Maxwell equations are linear equations with respect to the EM fields but
combining solutions for different $\epsilon$'s is like
adding apples and oranges.

Of course, this general argument applies to the Schr\"odinger equation as well.
For a particle moving in a potential, we have
\begin{eqnarray}
i\hbar\frac{\partial\Psi(\mathbf{r},t)}{\partial t}&=&
\left(\frac{1}{m}\mathbf{p}^2+V(\mathbf{r})\right)\Psi(\mathbf{r},t)
.
\label{schr0}
\end{eqnarray}
In essence, the WH Eq.~(\ref{three0}) asserts that
there is a relation between the solutions of
four problems defined by the potential $V(\mathbf{r})$ and three other potentials $V_j(\mathbf{r})$ for $j=1,2,3$.
More specifically, it asserts that
\begin{eqnarray}
\Psi(\mathbf{r},t)&=&
\Psi_1(\mathbf{r},t)+
\Psi_2(\mathbf{r},t)+
\Psi_3(\mathbf{r},t)
,
\label{schr1}
\end{eqnarray}
and
\begin{eqnarray}
V(\mathbf{r}) \Psi(\mathbf{r},t)&=&
V(\mathbf{r})
\Psi_1(\mathbf{r},t)+
V(\mathbf{r})
\Psi_2(\mathbf{r},t)+
V(\mathbf{r})\Psi_3(\mathbf{r},t)
\nonumber \\
&=&
V_1(\mathbf{r})\Psi_1(\mathbf{r},t)+
V_2(\mathbf{r})\Psi_2(\mathbf{r},t)+
V_3(\mathbf{r})\Psi_3(\mathbf{r},t)
.
\label{schr2}
\end{eqnarray}
The authors could not think of a general physical situation that would result in Eq.~(\ref{schr2}).

\section{Does the experiment reported in Ref.~\onlinecite{SINH10} have merit?}\label{merit}

Although in general, the WH of Refs.~\cite{SORK94,FRAN10,SINH10} does not hold,
the three-slit experiment reported in Ref.~\cite{SINH10} is not without merit,
as we now show by discussing FDTD simulations of an idealization of the
device used in the experiment~\cite{SINH10}.

\begin{figure*}[t]
\begin{center}
\mbox{
\includegraphics[width=8cm ]{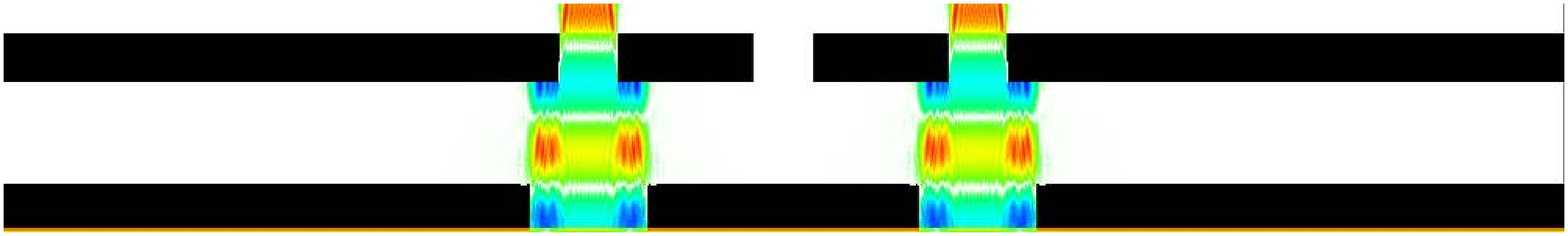}
\hbox to 0.5cm{}
\includegraphics[width=8cm ]{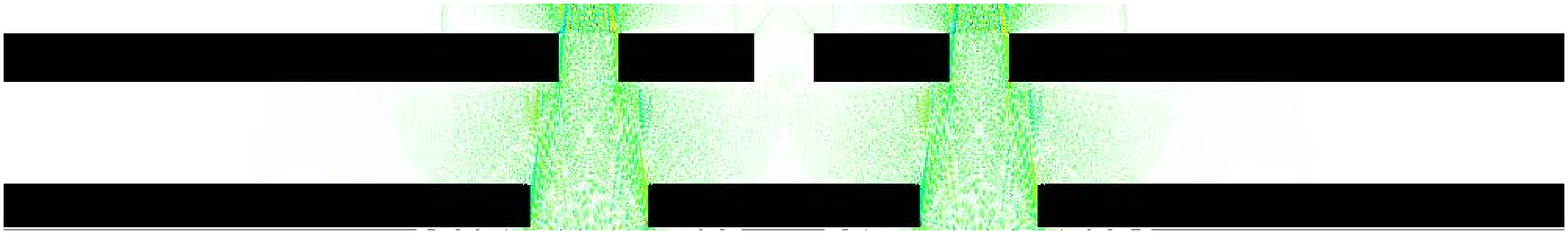}
}
\caption{Two-dimensional representation of the experiment reported in Ref.~\cite{SINH10}.
The three slits at the top are $30\mu$m wide, their centres being separated by $100\mu$m.
The blocking mask at the bottom can have one, two or three slits, each
slit being $60\mu$m wide with its centre
aligned with one of the slits in the top plate~\cite{SINH10}.
In the example shown, the middle slit of the blocking mask is closed (corresponding to the case OCO).
The separation between the top plate and blocking mask is $50\mu$m.
The index of refraction of the $25\mu$m-thick material (colored black) is $2.29+2.61 i$ (index
of refraction of iron at $405$nm).
The wavelength of the incident light is $405$nm.
Also shown are the amplitudes of the $E_x$ (left) and $E_z$ (right) components of the electric fields as obtained from
a FDTD solution of Maxwell's equation for a monochromatic light source (not shown) illuminating the blocking mask.
Note that the $E_x$- and $E_z$-components propagate in a very different manner.
}
\label{fig.greg0}
\end{center}
\end{figure*}

The geometry of this device is depicted in Fig.~\ref{fig.greg0} (see also Ref.~\onlinecite{SINH10}),
together with the stationary state FDTD solution of the Maxwell equations.
In the simulation, the device is illuminated from the bottom (Fig.~\ref{fig.lambda}), using a current source
that generates a monochromatic plane wave that propagates in the vertical direction.
The wavelength of the light, the dimension of the slits and their separation, blocking masks and material properties
are taken from Ref.~\cite{SINH10}.
In view of the large (compared to wavelength) dimensions of the slits, to reduce the computational burden,
we assume translational invariance in the direction along the long axis of the slits.
This idealization of the real experiment does not affect the conclusions, on the contrary: It
eliminates effects of the finite length of the slits.
In all these simulations, the 81 mesh points per wavelength ($\lambda=405$ nm) were taken to ensure that
the discretization errors of the EM fields and geometry are negligible.
The simulation box of $820\mu\mathrm{m}\times120\mu\mathrm{m}$
(corresponding to 3 936 188 001 grid points) contains UPML layers
to eliminate reflection from the boundaries~\cite{TAFL05}.
Each calculation requires about 900GB of memory and took about 12 hours,
using 8192 processors of the IBM BlueGene/P at the J\"ulich Supercomputing Centre.

\begin{figure*}[ht]
\begin{center}
\includegraphics[width=7cm ]{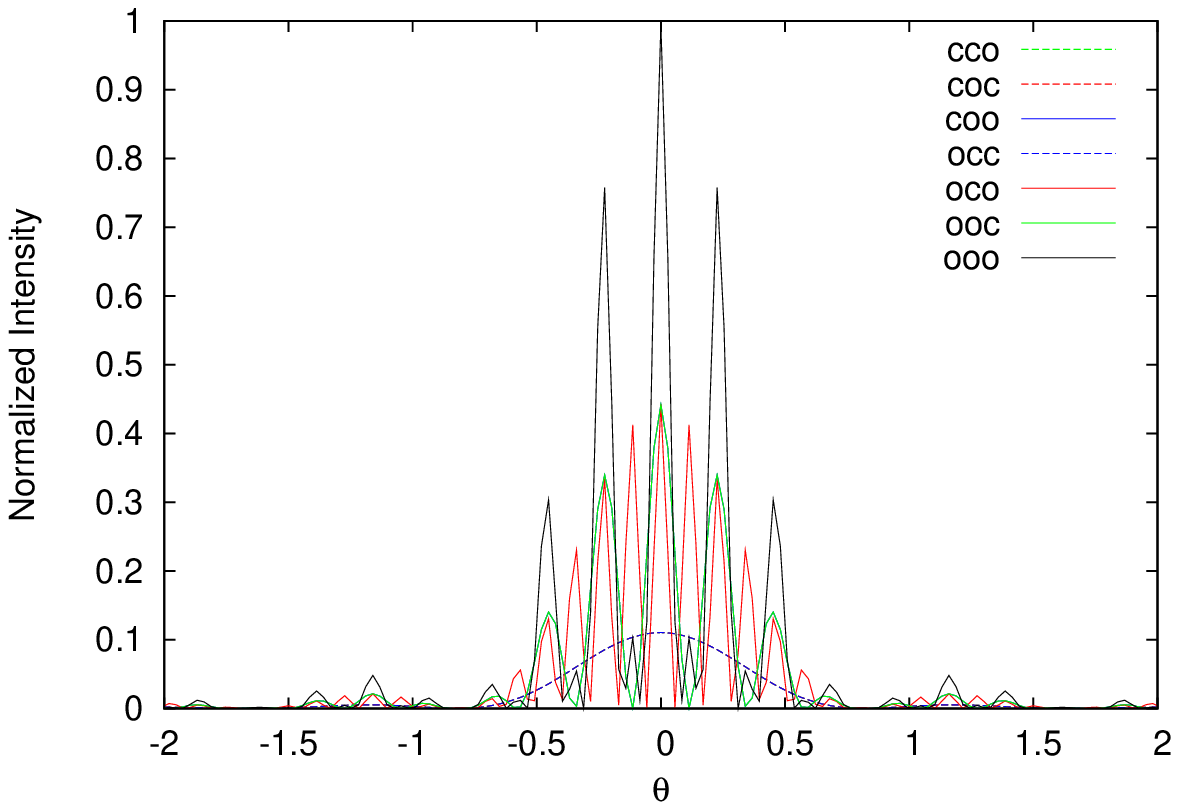}
\includegraphics[width=7cm ]{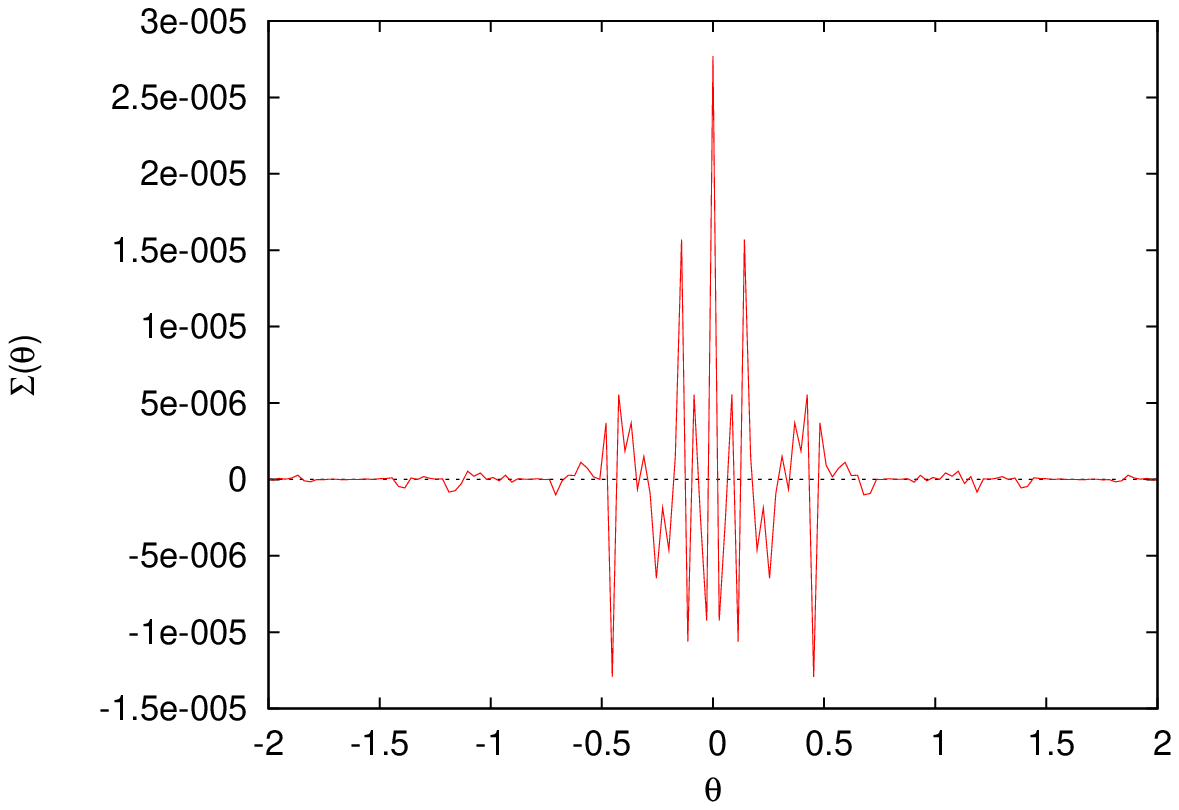}
\caption{Left:
Angular distribution of light transmitted by the system shown in Fig.~\ref{fig.greg0}
for the cases in which all slits are open (OOO), one slit is closed (COO,OCO,OOC),
and two slits are closed (CCO,COC,OCC), as obtained from FDTD simulations.
Right: Normalized difference
$\Sigma(\theta)=[I(\theta;\mathrm{OOO})-I(\theta;\mathrm{COO})-I(\theta;\mathrm{OCO})-I(\theta;\mathrm{OOC})+I(\theta;\mathrm{CCO})
+I(\theta;\mathrm{COC})+I(\theta;\mathrm{OCC})]/I(\theta=0;\mathrm{OOO})$
as a function of $\theta$.
According to the WH~\cite{SORK94,SINH10}, this difference should be zero.
Note that the symmetric structure of $\Sigma(\theta)$ rules out that the signal
is due to numerical noise.
}
\label{fig.greg0.combine}
\end{center}
\end{figure*}

Qualitatively, Fig.~\ref{fig.greg0}(left) indicates that the $x$-component ($E_x$) of the EM-field
propagates through the two layers of slits with very little diffraction from the top (=blocking) layer.
This is not the case for the $z$-component shown in Fig.~\ref{fig.greg0}(right).
In this idealized simulation setup, the amplitude of the $y$-component of the EM-field is zero.

In Fig.~\ref{fig.greg0.combine}(left) we present the results for the angular distribution
of the seven cases (OOO, OOC, OCO, COO, OCC, COC, and CCO), extracted from seven FDTD simulations
(curves with two slits closed overlap).
From Fig.~\ref{fig.greg0.combine}(right), it is clear that $\Sigma(\theta)$ is not identically zero.
However, $\Sigma(\theta)$ is of the order of $10^{-5}$
and such small values of $\Sigma(\theta)$ are unlikely to be observable in an
experiment such as the one reported in Ref.~\onlinecite{SINH10}.
Another way of phrasing this is to say that the experiment~\cite{SINH10} is performed
in a regime in which scalar Fraunhofer theory works well,
as can be expected from the dimensions of the slits and slit separations of the device.
As we have shown in Section~\ref{whholds}, it is precisely in this limit
that the WH holds.
Therefore, assuming that the non-linearity of the detectors used in the experiment~\cite{SINH10}
is negligible, the experiment~\cite{SINH10} suggests that
the measured intensity is proportional to the square of the amplitudes
of the wave field interacting with the detector.

\section{Discussion}\label{discussion}

A necessary condition for a mathematical model to give a logically consistent description of the experimental facts
is that there is one-to-one correspondence between the symbols in the mathematical description
and the actual experimental configurations.
When applied to the three-slit experiment in which one slit or two slits may be closed,
the argument that leads from Eq.~(\ref{three0}) to Eq.~(\ref{three1}) is fundamentally flawed
in that there is no such correspondence.

If $\psi_j$ in Eq.~(\ref{three0}) is to represent the amplitude of the wave emanating from the $j$th slit with
all other slits closed, the WH should be written as
\begin{eqnarray}
I(\mathbf{r},\mathrm{OOO})&=&|\psi_1(\mathbf{r},\mathrm{OCC})+\psi_2(\mathbf{r},\mathrm{COC})+\psi_3(\mathbf{r},\mathrm{CCO})|^2
,
\label{three0x}
\end{eqnarray}
that is, we should label the $\psi's$ such that there can be no doubt about the experiment that they describe.
This notation establishes the necessary one-to-one correspondence between the mathematical description (the $\psi$'s)
of the particular experiment (labeled by $\mathrm{OCC}$, etc.).
Now, we have
\begin{eqnarray}
I(\mathbf{r},\mathrm{OOO})
&=&|\psi_1(\mathbf{r},\mathrm{OCC})+\psi_2(\mathbf{r},\mathrm{COC})|^2
+|\psi_1(\mathbf{r},\mathrm{OCC})+\psi_3(\mathbf{r},\mathrm{CCO})|^2
\nonumber \\&&
+|\psi_2(\mathbf{r},\mathrm{COC})+\psi_3(\mathbf{r},\mathrm{CCO})|^2
\nonumber \\&&
-|\psi_1(\mathbf{r},\mathrm{OCC})|^2-|\psi_2(\mathbf{r},\mathrm{CCO})|^2-|\psi_3(\mathbf{r},\mathrm{CCO})|^2
.
\label{three3}
\end{eqnarray}
At this point, it is simply impossible to bring Eq.~(\ref{three3}) into the form Eq.~(\ref{three1})
without making the assumption that
\begin{eqnarray}
\psi(\mathbf{r},\mathrm{OOC}) &=& \psi_1(\mathbf{r},\mathrm{OCC})+\psi_2(\mathbf{r},\mathrm{COC}),
\nonumber \\
\psi(\mathbf{r},\mathrm{OCO}) &=& \psi_1(\mathbf{r},\mathrm{OCC})+\psi_3(\mathbf{r},\mathrm{CCO}),
\nonumber \\
\psi(\mathbf{r},\mathrm{COO}) &=& \psi_2(\mathbf{r},\mathrm{COC})+\psi_3(\mathbf{r},\mathrm{CCO}).
\label{three4}
\end{eqnarray}
If we accept this assumption, we recover Eq.~(\ref{three1}).
However, the assumption expressed by Eq.~(\ref{three4}) cannot be justified from general principles of quantum theory
or Maxwell's theory: The only way to ``justify'' Eq.~(\ref{three4}) is to ``forget'' that
the $\psi$'s are labeled by the type of experiment (e.g. $\mathrm{OCC}$) they describe.
For a discussion of this point in the case of a two-slit experiment, see Refs.~\onlinecite{BALL86,BALL03}.

In other words, accepting Eq.~(\ref{three4}) destroys the one-to-one correspondence between
the symbols in the mathematical theory and the different experimental configurations,
opening the route to conclusions that cannot be derived from the theory proper.
Hence, if $\Delta(\mathbf{r})$ would indeed be different from zero
for a three-slit experiment, one cannot conclude that Born's rule does not strictly hold.
However, if $\Delta(\mathbf{r})$ is found to be nonzero in the case of the Mach-Zehnder interferometer experiment described
in Ref.~\onlinecite{FRAN10} and non-linearity of the apparatus can be ruled out, the conclusion would
be that Born's rule does not hold.

\section{Summary}\label{summary}
The results of this paper can be summarized as follows:
\begin{enumerate}
\item{The third-order interference, as obtained from explicit solutions of Maxwell's equations
for realistic models of three-slit devices, is nonzero.}
\item{
The hypothesis~\cite{SORK94,FRAN10,SINH10} that the third-order interference Eq.~(\ref{three2})
should be zero is fatally
flawed because it requires dropping the one-to-one correspondence
between the symbols in the mathematical theory and the different experimental configurations.
}
\item{Having shown that the expression for the third-order interference Eq.~(\ref{three2}) cannot be derived from
quantum theory or Maxwell theory of a three-slit experiment, it follows that any conclusion based on this expression
has nothing to say about these theories (with the exceptions mentioned in Section~\ref{whholds}).}
\end{enumerate}

\section{Acknowledgement}
This work is partially supported by NCF, the Netherlands.

\bibliography{c:/d/papers/epr11}   

\begin{thebibliography}{15}
\expandafter\ifx\csname natexlab\endcsname\relax\def\natexlab#1{#1}\fi
\expandafter\ifx\csname bibnamefont\endcsname\relax
  \def\bibnamefont#1{#1}\fi
\expandafter\ifx\csname bibfnamefont\endcsname\relax
  \def\bibfnamefont#1{#1}\fi
\expandafter\ifx\csname citenamefont\endcsname\relax
  \def\citenamefont#1{#1}\fi
\expandafter\ifx\csname url\endcsname\relax
  \def\url#1{\texttt{#1}}\fi
\expandafter\ifx\csname urlprefix\endcsname\relax\def\urlprefix{URL }\fi
\providecommand{\bibinfo}[2]{#2}
\providecommand{\eprint}[2][]{\url{#2}}

\bibitem[{\citenamefont{Sorkin}(1994)}]{SORK94}
\bibinfo{author}{\bibfnamefont{R.~D.} \bibnamefont{Sorkin}},
  \bibinfo{journal}{Mod. Phys. Lett.} \textbf{\bibinfo{volume}{9}},
  \bibinfo{pages}{3119 } (\bibinfo{year}{1994}).

\bibitem[{\citenamefont{Franson}(2010)}]{FRAN10}
\bibinfo{author}{\bibfnamefont{J.}~\bibnamefont{Franson}},
  \bibinfo{journal}{Science} \textbf{\bibinfo{volume}{329}},
  \bibinfo{pages}{396 } (\bibinfo{year}{2010}).

\bibitem[{\citenamefont{Sinha et~al.}(2010)\citenamefont{Sinha, Couteau,
  Jennewein, Laflamme, and G.Weihs}}]{SINH10}
\bibinfo{author}{\bibfnamefont{U.}~\bibnamefont{Sinha}},
  \bibinfo{author}{\bibfnamefont{C.}~\bibnamefont{Couteau}},
  \bibinfo{author}{\bibfnamefont{T.}~\bibnamefont{Jennewein}},
  \bibinfo{author}{\bibfnamefont{R.}~\bibnamefont{Laflamme}}, \bibnamefont{and}
  \bibinfo{author}{\bibnamefont{G.Weihs}}, \bibinfo{journal}{Science}
  \textbf{\bibinfo{volume}{329}}, \bibinfo{pages}{418 } (\bibinfo{year}{2010}).

\bibitem[{\citenamefont{Taflove and Hagness}(2005)}]{TAFL05}
\bibinfo{author}{\bibfnamefont{A.}~\bibnamefont{Taflove}} \bibnamefont{and}
  \bibinfo{author}{\bibfnamefont{S.}~\bibnamefont{Hagness}},
  \emph{\bibinfo{title}{Computational Electrodynamics: The Finite-Difference
  Time-Domain Method}} (\bibinfo{publisher}{Artech House},
  \bibinfo{address}{Boston}, \bibinfo{year}{2005}).

\bibitem[{\citenamefont{Born and Wolf}(1964)}]{BORN64}
\bibinfo{author}{\bibfnamefont{M.}~\bibnamefont{Born}} \bibnamefont{and}
  \bibinfo{author}{\bibfnamefont{E.}~\bibnamefont{Wolf}},
  \emph{\bibinfo{title}{{Principles of Optics}}}
  (\bibinfo{publisher}{Pergamon}, \bibinfo{address}{Oxford},
  \bibinfo{year}{1964}).

\bibitem[{\citenamefont{Gay et~al.}(2006)\citenamefont{Gay, Alloschery, {Viaris
  de Lesegno}, {O'Dwyer}, Weiner, and Lezec}}]{GAY06}
\bibinfo{author}{\bibfnamefont{G.}~\bibnamefont{Gay}},
  \bibinfo{author}{\bibfnamefont{O.}~\bibnamefont{Alloschery}},
  \bibinfo{author}{\bibfnamefont{B.}~\bibnamefont{{Viaris de Lesegno}}},
  \bibinfo{author}{\bibfnamefont{C.}~\bibnamefont{{O'Dwyer}}},
  \bibinfo{author}{\bibfnamefont{J.}~\bibnamefont{Weiner}}, \bibnamefont{and}
  \bibinfo{author}{\bibfnamefont{H.~J.} \bibnamefont{Lezec}},
  \bibinfo{journal}{Nature Phys.} \textbf{\bibinfo{volume}{2}},
  \bibinfo{pages}{262 } (\bibinfo{year}{2006}).

\bibitem[{\citenamefont{Lalanne and Hugonin}(2006)}]{LALA06}
\bibinfo{author}{\bibfnamefont{P.}~\bibnamefont{Lalanne}} \bibnamefont{and}
  \bibinfo{author}{\bibfnamefont{J.~P.} \bibnamefont{Hugonin}},
  \bibinfo{journal}{Nature Phys.} \textbf{\bibinfo{volume}{2}},
  \bibinfo{pages}{551 } (\bibinfo{year}{2006}).

\bibitem[{\citenamefont{Roychoudhuri}(2010{\natexlab{a}})}]{ROYC10}
\bibinfo{author}{\bibfnamefont{C.}~\bibnamefont{Roychoudhuri}}, in
  \emph{\bibinfo{booktitle}{Quantum Theory: Reconsideration of Foundations -
  5}}, edited by \bibinfo{editor}{\bibfnamefont{A.}~\bibnamefont{Khrennikov}}
  (\bibinfo{publisher}{AIP Conference Proceedings}, \bibinfo{address}{Melville
  and New York}, \bibinfo{year}{2010}{\natexlab{a}}), vol.
  \bibinfo{volume}{1232}, pp. \bibinfo{pages}{143 -- 152}.

\bibitem[{\citenamefont{Roychoudhuri}(2010{\natexlab{b}})}]{ROYC10b}
\bibinfo{author}{\bibfnamefont{C.}~\bibnamefont{Roychoudhuri}},
  \bibinfo{journal}{J. Nanophotonics} \textbf{\bibinfo{volume}{4}},
  \bibinfo{pages}{043512} (\bibinfo{year}{2010}{\natexlab{b}}).

\bibitem[{\citenamefont{Feynman and Hibbs}(1965)}]{FEYN65b}
\bibinfo{author}{\bibfnamefont{R.~P.} \bibnamefont{Feynman}} \bibnamefont{and}
  \bibinfo{author}{\bibfnamefont{A.~R.} \bibnamefont{Hibbs}},
  \emph{\bibinfo{title}{Quantum Mechanics and Path Integrals}}
  (\bibinfo{publisher}{McGraw-Hill}, \bibinfo{address}{New York},
  \bibinfo{year}{1965}).

\bibitem[{\citenamefont{Khrennikov}(2008)}]{KHRE08a}
\bibinfo{author}{\bibfnamefont{A.}~\bibnamefont{Khrennikov}},
  \bibinfo{journal}{Phys. Lett. A} \textbf{\bibinfo{volume}{372}},
  \bibinfo{pages}{6588 } (\bibinfo{year}{2008}),
  \bibinfo{note}{{\url{http://arxiv.org/abs/0805.1511v3}}}.

\bibitem[{\citenamefont{Ududec et~al.}(2010)\citenamefont{Ududec, Barnum, and
  Emerson}}]{UDUD10}
\bibinfo{author}{\bibfnamefont{C.}~\bibnamefont{Ududec}},
  \bibinfo{author}{\bibfnamefont{H.}~\bibnamefont{Barnum}}, \bibnamefont{and}
  \bibinfo{author}{\bibfnamefont{J.}~\bibnamefont{Emerson}},
  \bibinfo{journal}{Found. Phys.} \textbf{\bibinfo{volume}{{ }}},
  \bibinfo{pages}{DOI: 10.1007/s10701} (\bibinfo{year}{2010}).

\bibitem[{\citenamefont{Nyman and Basieva}(2011)}]{NYMA11}
\bibinfo{author}{\bibfnamefont{P.}~\bibnamefont{Nyman}} \bibnamefont{and}
  \bibinfo{author}{\bibfnamefont{I.}~\bibnamefont{Basieva}}, in
  \emph{\bibinfo{booktitle}{Advances in Quantum Theory}}, edited by
  \bibinfo{editor}{\bibfnamefont{G.}~\bibnamefont{Jaeger}},
  \bibinfo{editor}{\bibfnamefont{A.}~\bibnamefont{Khrennikov}},
  \bibinfo{editor}{\bibfnamefont{M.}~\bibnamefont{Schlosshauer}},
  \bibnamefont{and} \bibinfo{editor}{\bibfnamefont{G.}~\bibnamefont{Weihs}}
  (\bibinfo{publisher}{AIP Conference Proceedings}, \bibinfo{address}{Melville,
  New York}, \bibinfo{year}{2011}), vol. \bibinfo{volume}{1327}, p.
  \bibinfo{pages}{439}, \eprint{{\url{http://arxiv.org/abs/1010.1142}}}.

\bibitem[{\citenamefont{Ballentine}(1986)}]{BALL86}
\bibinfo{author}{\bibfnamefont{L.~E.} \bibnamefont{Ballentine}},
  \bibinfo{journal}{Am. J. Phys.} \textbf{\bibinfo{volume}{54}},
  \bibinfo{pages}{883 } (\bibinfo{year}{1986}).

\bibitem[{\citenamefont{Ballentine}(2003)}]{BALL03}
\bibinfo{author}{\bibfnamefont{L.~E.} \bibnamefont{Ballentine}},
  \emph{\bibinfo{title}{{Quantum Mechanics: A Modern Development}}}
  (\bibinfo{publisher}{World Scientific}, \bibinfo{address}{Singapore},
  \bibinfo{year}{2003}).

\end{thebibliography}

\end{document}